\documentclass[conference]{IEEEtran}
\usepackage[compatibility=false]{caption}
\makeatletter

\def\ps@IEEEtitlepagestyle{%
  \def\@oddfoot{\mycopyrightnotice}%
  \def\@evenfoot{}%
}
\def\mycopyrightnotice{%
  {\footnotesize \hfill}
  \gdef\mycopyrightnotice{}
}

\usepackage{blindtext}
\usepackage{eso-pic}
\IEEEoverridecommandlockouts
\usepackage{cite}
\usepackage{amsmath,amssymb,amsfonts}
\usepackage{algorithmic}
\usepackage{graphicx}
\usepackage{textcomp}
\usepackage{xcolor}
\def\BibTeX{{\rm B\kern-.05em{\sc i\kern-.025em b}\kern-.08em
    T\kern-.1667em\lower.7ex\hbox{E}\kern-.125emX}}
    
\usepackage{eso-pic}
\newcommand\AtPageUpperMyright[1]{\AtPageUpperLeft{%
 \put(\LenToUnit{0.17\paperwidth},\LenToUnit{-2cm}){%
     \parbox{0.9\textwidth}{\raggedleft\fontsize{8}{11}\selectfont #1}}%
 }}%
\newcommand{\conf}[1]{%
\AddToShipoutPictureBG*{%
\AtPageUpperMyright{#1}
}
}    

\usepackage{graphicx}
\usepackage[labelfont=bf,compatibility=false]{caption}
\usepackage{url}
\usepackage{float}
\usepackage{hyperref}
\usepackage{amsmath}
 
\usepackage{mathtools}
\usepackage{bm, csquotes}
\usepackage{subcaption}
\usepackage{chngcntr}
\counterwithin{figure}{section}
\usepackage{multirow}
\usepackage{array}
\usepackage{enumerate}
\usepackage{tikz}
\usetikzlibrary{arrows.meta, positioning, shapes.geometric}
\usepackage{booktabs}

\begin{document}
\title{\vspace*{1cm} Unsupervised Detection of Groundwater Storage Anomalies in Ghana Using GRACE Satellite Data\\
}

\author{\IEEEauthorblockN{George Yamoah Afrifa}
\IEEEauthorblockA{Ghana Space Science and\\ Technology Institute\\
Accra, Ghana\\
george.afrifa@gaec.gov.gh}
\and
\IEEEauthorblockN{Theophilus Ansah-Narh}
\IEEEauthorblockA{Ghana Space Science and\\ Technology Institute\\
Accra, Ghana\\
theophilus.ansah-narh@gaec.gov.gh}
\and
\IEEEauthorblockN{Marcellin Atemkeng\textsuperscript{$\dagger$}}
\IEEEauthorblockA{Department of Mathematics\\Rhodes University, Grahamstown, South Africa\\
m.atemkeng@ru.ac.za}
}

\maketitle
\conf{\textit{  Proc. of the International Conference on Electrical, Computer, Communications and Mechatronics Engineering (ICECCME 2026) \\ 
15-17 October 2026, Bali, Indonesia}}

\begin{abstract}
   Groundwater variability in Ghana remains poorly characterized due to limited long-term in-situ observations. This study investigates groundwater storage anomalies using GRACE-derived data from 2004--2024 combined with statistical analysis and unsupervised machine learning. Groundwater anomalies were standardized using Z-scores, while an ensemble-based Isolation Forest framework was applied for anomaly detection. The results revealed substantial temporal variability, with persistent groundwater deficits during 2004--2009 followed by increasing positive anomalies after 2018. A total of 12 anomalous months were identified, comprising 5 deficit and 7 surplus events, with the strongest anomalies associated with groundwater deficits. Spatial analysis showed more frequent deficit anomalies in northern Ghana and stronger surplus occurrence in southern regions. Comparison with statistical thresholds further indicated that the machine learning framework captured additional subtle deviations beyond conventional threshold-based methods. Overall, the integration of GRACE observations with unsupervised anomaly detection provides a practical framework for groundwater monitoring in data-scarce environments.
\end{abstract}



\section{Introduction} \label{intro}

Groundwater is a critical component of the hydrological cycle and an important source of freshwater for domestic, agricultural, and industrial activities, particularly in data-scarce regions such as sub-Saharan Africa, where it supports livelihoods and enhances resilience to climate variability \cite{macdonald2021mapping}. In Ghana, groundwater serves as a major source of potable water, especially in rural communities that depend heavily on boreholes and shallow aquifers \cite{obuobie2012groundwater}. However, effective monitoring of groundwater systems remains challenging because of sparse in-situ observations and limited long-term hydrological records.

The Gravity Recovery and Climate Experiment (GRACE) and GRACE Follow-On (GRACE-FO) satellite missions have significantly improved large-scale monitoring of terrestrial water storage (TWS), enabling the estimation of groundwater storage variations after removing other hydrological components such as soil moisture and surface water \cite{liu2025groundwater}. GRACE-derived groundwater datasets have been widely used to investigate groundwater depletion and variability in regions including India, California, and the Sahel \cite{famiglietti2011satellites}. Most existing studies, however, primarily employ conventional approaches such as trend analysis and standardized anomaly indices, which may not adequately capture subtle or nonlinear variations in complex hydrological systems.

Recent advances in machine learning have introduced unsupervised anomaly detection techniques capable of identifying unusual patterns in environmental time series without requiring labeled datasets \cite{holden2013effective}. Among these methods, the Isolation Forest algorithm has gained attention because of its computational efficiency and effectiveness in isolating anomalous observations through random partitioning \cite{al2021isolation}. The method has shown promising performance in hydrological and environmental applications \cite{qin2019hydrological}. However, its application to GRACE-derived groundwater storage datasets remains limited, particularly in West Africa, where previous studies have largely focused on groundwater trends and variability analysis \cite{bonsor2018seasonal}.

Unlike previous GRACE-based groundwater studies in West Africa that mainly employ trend analysis and standardized anomaly indices, this study integrates an Isolation Forest-based framework to detect subtle and nonlinear groundwater anomalies in Ghana. The framework is applied to groundwater storage observations derived from GRACE and GRACE-FO data for the period 2004--2024, and the detected anomalies are evaluated against conventional standardized anomaly metrics. The objective is to improve understanding of the spatiotemporal variability of groundwater storage and provide a data-driven framework for groundwater anomaly monitoring in data-limited environments.

\section{Study Area and Data Description} \label{sec:EDA}
\subsection{Study Area}

\begin{figure*}
	\begin{minipage}{\linewidth}
		\centering
		\includegraphics[width=\textwidth]{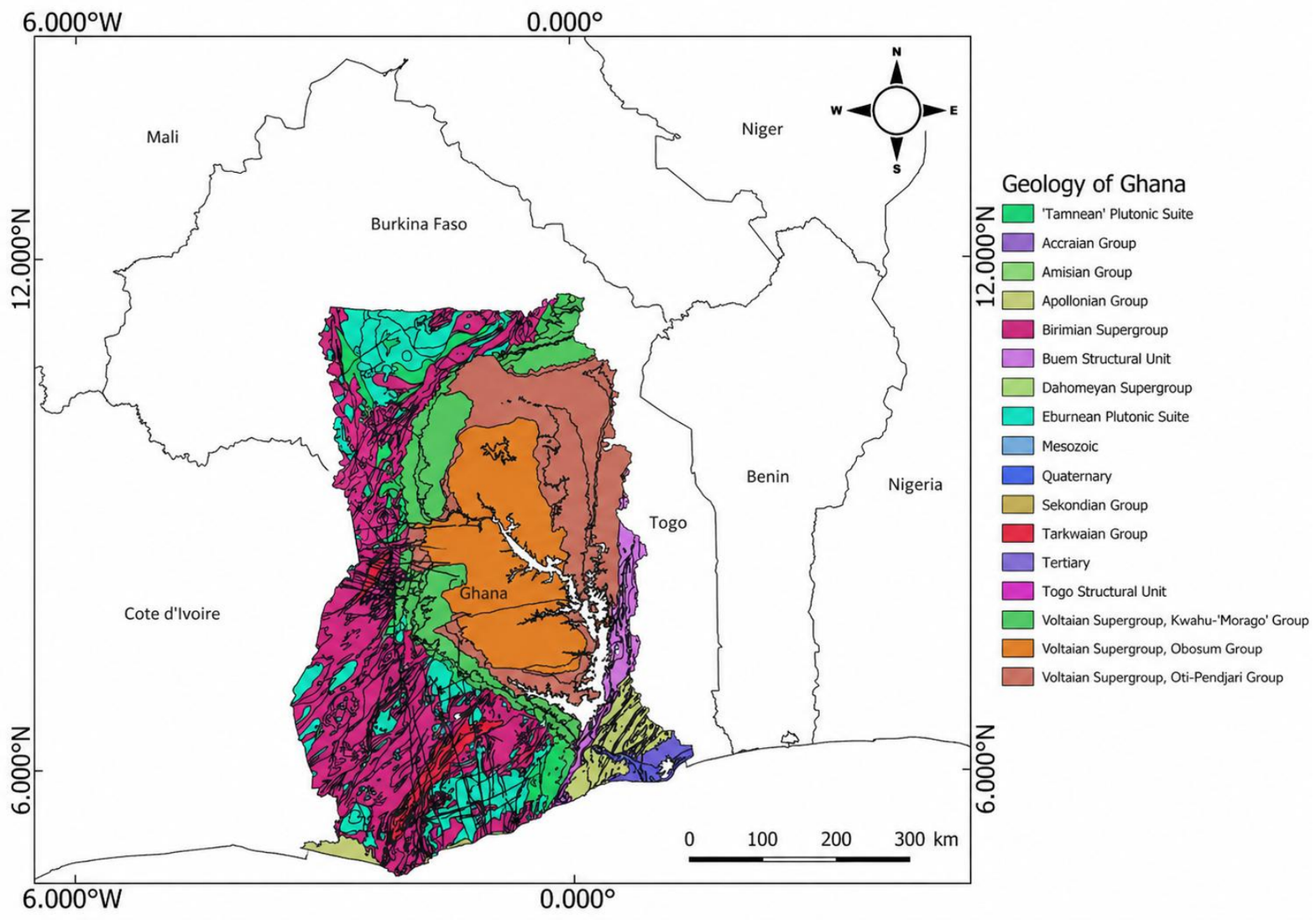} %
	\end{minipage}
	\caption{Geological map of Ghana showing the spatial distribution of major lithological units across the country. The map also highlights national boundaries, neighbouring countries, and key geographic references, providing geological context for groundwater storage variability analysis.}
	\label{fig:GH_geology1}
\end{figure*}
Ghana is a West African country located between latitudes $4^\circ44'$N and $11^\circ11'$N and longitudes $1^\circ12'$E and $3^\circ11'$W, covering approximately $238{,}535$~km$^2$. It is bordered by C\^ote d'Ivoire, Burkina Faso, Togo, and the Gulf of Guinea. The country exhibits diverse climatic, geological, and hydrological conditions, making it suitable for groundwater variability studies. Ghana’s climate is tropical and influenced by the Inter-Tropical Convergence Zone (ITCZ), resulting in bimodal rainfall in the south (March--July and September--November) and unimodal rainfall in the north (May--October) \cite{amekudzi2015variabilities}. Annual precipitation ranges from about 800~mm in the coastal savannah to over 2{,}000~mm in the southwestern forest zone, strongly controlling groundwater recharge processes \cite{obuobie2012groundwater}. High evapotranspiration rates, often exceeding precipitation during the dry season, further constrain recharge \cite{acheampong1988water}. Hydrologically, Ghana is dominated by the Volta River Basin, which covers about 70\% of the country and includes major rivers and Lake Volta, necessitating reliance on groundwater in areas where surface water is variable \cite{agyemang2025review}. 
The hydrogeology consists of three main formations: the Precambrian crystalline basement complex (about 54\%), the Voltaian sedimentary basin (about 45\%), and coastal sedimentary formations. Groundwater in the basement complex occurs in weathered and fractured zones with low storage capacity, while the Voltaian Basin exhibits low permeability and discontinuous aquifers \cite{obuobie2012groundwater}. Coastal sedimentary aquifers are generally more productive but vulnerable to over-extraction and saline intrusion \cite{banoeng2010hydrogeology}. 
Groundwater supplies approximately 60--70\% of potable water in Ghana, particularly in rural areas, and supports domestic, agricultural, and industrial uses \cite{macdonald2021mapping}. However, groundwater systems face challenges such as over-abstraction, contamination, and climate variability, with evidence of declining levels in some regions \cite{cuthbert2019global}. The limited availability of long-term in-situ monitoring data further constrains comprehensive assessment. The spatial scale of the Volta Basin aligns well with the resolution of GRACE observations ($\sim$300 --400~km), enabling effective analysis of terrestrial water storage variations and groundwater dynamics \cite{bonsor2018seasonal}.
Given its climatic variability, hydrogeological diversity, and dependence on groundwater resources, Ghana provides an appropriate and representative case study for applying unsupervised anomaly detection techniques to GRACE-derived groundwater storage data.

\subsection{Data Description and Groundwater Anomaly Computation}

Groundwater storage data used in this study were derived from GRACE/GRACE-FO terrestrial water storage observations covering the period 2004--2024. Monthly groundwater storage values were analyzed over spatial grid cells defined by latitude and longitude coordinates across Ghana. The time variable, originally expressed in decimal years, was converted to a standard datetime format to facilitate temporal analysis, after which the dataset was organized into a spatiotemporal framework based on individual grid cells.

Groundwater storage ($\rm GWS$) anomalies were estimated from GRACE terrestrial water storage ($\rm TWS$) observations after removing other hydrological components, including soil moisture, surface water, snow water equivalent, and canopy water. The groundwater storage formulation is expressed as:

\begin{equation}
GWS = TWS - (SM + SW + SWE + CW),
\end{equation}

\noindent  following standard GRACE-based groundwater estimation approaches \cite{zhang2022bridging}.
where $\rm SM$ represents soil moisture, $\rm SW$ denotes surface water, $\rm SWE$ is snow water equivalent, and $\rm CW$ represents canopy water. In the context of Ghana, snow-related components are negligible.

To quantify deviations from long-term groundwater conditions, groundwater storage anomalies ($\rm GWSA$) were computed for each grid cell relative to the temporal mean:

\begin{equation}
GWSA_{i,t} = GWS_{i,t} - \overline{GWS}_{i},
\end{equation}

\noindent  where $GWS_{i,t}$ denotes groundwater storage at grid cell $i$ and time $t$, while $\overline{GWS}_{i}$ represents the long-term mean groundwater storage at grid cell $i$.

To enable comparison across locations with different variability characteristics, anomalies were standardized using the Z-score:

\begin{equation}
Z_{i,t} = \frac{GWS_{i,t} - \overline{GWS}_{i}}{\sigma_i},
\end{equation}

\noindent  where $\sigma_i$ is the standard deviation of groundwater storage at grid cell $i$. Thresholds of $\pm1$ and $\pm2$ were used to indicate moderate and extreme anomalies, respectively.

For national-scale analysis, groundwater anomalies were spatially aggregated by computing the mean anomaly across all grid cells within Ghana at each time step:

\begin{equation}
\overline{GWSA}_t = \frac{1}{N} \sum_{i=1}^{N} GWSA_{i,t},
\end{equation}

\noindent  where $N$ represents the total number of grid cells within the study area. This aggregation produced a country-scale time series describing the temporal evolution of groundwater variability in Ghana.

\section{Anomaly Detection Techniques} \label{sec:SE}

This study employed an unsupervised machine learning framework to investigate anomalous groundwater storage behavior in Ghana using GRACE-derived groundwater storage anomalies. The methodology integrated statistical anomaly characterization, ensemble-based anomaly detection, and spatial analysis to examine the temporal and spatial variability of groundwater dynamics.

\subsection{Unsupervised Anomaly Detection}

Anomaly detection was performed using an ensemble isolation-based learning approach designed to identify rare observations through recursive random partitioning of the feature space \cite{al2021isolation,xu2023deep}. The method operates on the principle that anomalous observations are statistically sparse and therefore require fewer partitions to become isolated compared to normal observations embedded within dense regions of the data distribution.

This approach is particularly suitable for groundwater systems because hydroclimatic processes often exhibit nonlinear variability, nonstationary temporal behavior, and irregular extreme events that are difficult to characterize using conventional threshold-based methods. In contrast to supervised learning techniques, the framework does not require predefined labels or assumptions regarding the statistical distribution of anomalies.

For each spatial grid cell, the $\rm GWSA$ time series was independently analyzed. Let

\begin{equation}
\mathbf{x}_i = \{GWSA_{i,1}, GWSA_{i,2}, \ldots, GWSA_{i,T}\}
\end{equation}

\noindent represent the temporal anomaly sequence for grid cell $i$, where $T$ denotes the total number of monthly observations. Each sequence was reshaped into a one-dimensional feature vector prior to model training.

The ensemble learning procedure constructs multiple binary isolation trees by recursively selecting random partition thresholds within the feature space. For a given observation $x$, the number of recursive partitions required to isolate the observation is referred to as the path length, denoted by $h(x)$. Observations associated with shorter average path lengths are considered more anomalous because they occupy isolated regions of the distribution.
The anomaly score is estimated as:

\begin{equation}
s(x,n) = 2^{-\frac{E(h(x))}{c(n)}},
\end{equation}

\noindent  where $E(h(x))$ denotes the expected path length of observation $x$ across the ensemble, $n$ is the sample size, and $c(n)$ represents a normalization factor corresponding to the average path length of binary search trees \cite{liu2012isolation}. Values of $s(x,n)$ approaching 1 indicate strong anomalous behavior, while values closer to 0 correspond to normal observations.

Model implementation was performed using the \texttt{scikit-learn} framework with a contamination parameter of $0.05$, corresponding to an expected anomaly proportion of $5\%$. A fixed random state of 42 was used to ensure reproducibility of the stochastic partitioning process. Monthly anomaly scores were subsequently computed for all valid groundwater storage observations across the study region.

An anomaly classification function was defined as:

\begin{equation}
A_t =
\begin{cases}
1, & \text{if } s_t > \tau \\
0, & \text{otherwise}
\end{cases}
\end{equation}

\noindent where $A_t$ denotes the anomaly state at time step $t$, $s_t$ represents the anomaly score associated with the observation, and $\tau$ is the anomaly threshold determined internally from the contamination parameter.

To identify periods of extreme groundwater behavior, monthly anomaly scores were ranked according to their magnitude, with highly ranked observations interpreted as significant deviations from the dominant groundwater variability regime. Because the detection framework is distribution-independent, both groundwater deficit and surplus anomalies could be identified without imposing predefined directional assumptions.

\subsection{Comparison with Statistical Thresholds}

To examine the consistency between data-driven anomaly detection and conventional statistical characterization, the detected anomalous periods were compared with anomalies identified using standardized Z-score thresholds. The comparison was intended to evaluate the degree of correspondence between threshold-based and ensemble-based anomaly identification rather than treating the statistical approach as absolute ground truth.

A confusion matrix was constructed to summarize shared detections, machine learning-only detections, and statistically identified anomalies. This comparison provided insight into the capability of the ensemble framework to identify subtle or nonlinear deviations that may not exceed fixed statistical thresholds.

\begin{figure*}
	\begin{minipage}{\linewidth}
		\centering
		\includegraphics[width=0.8\textwidth]{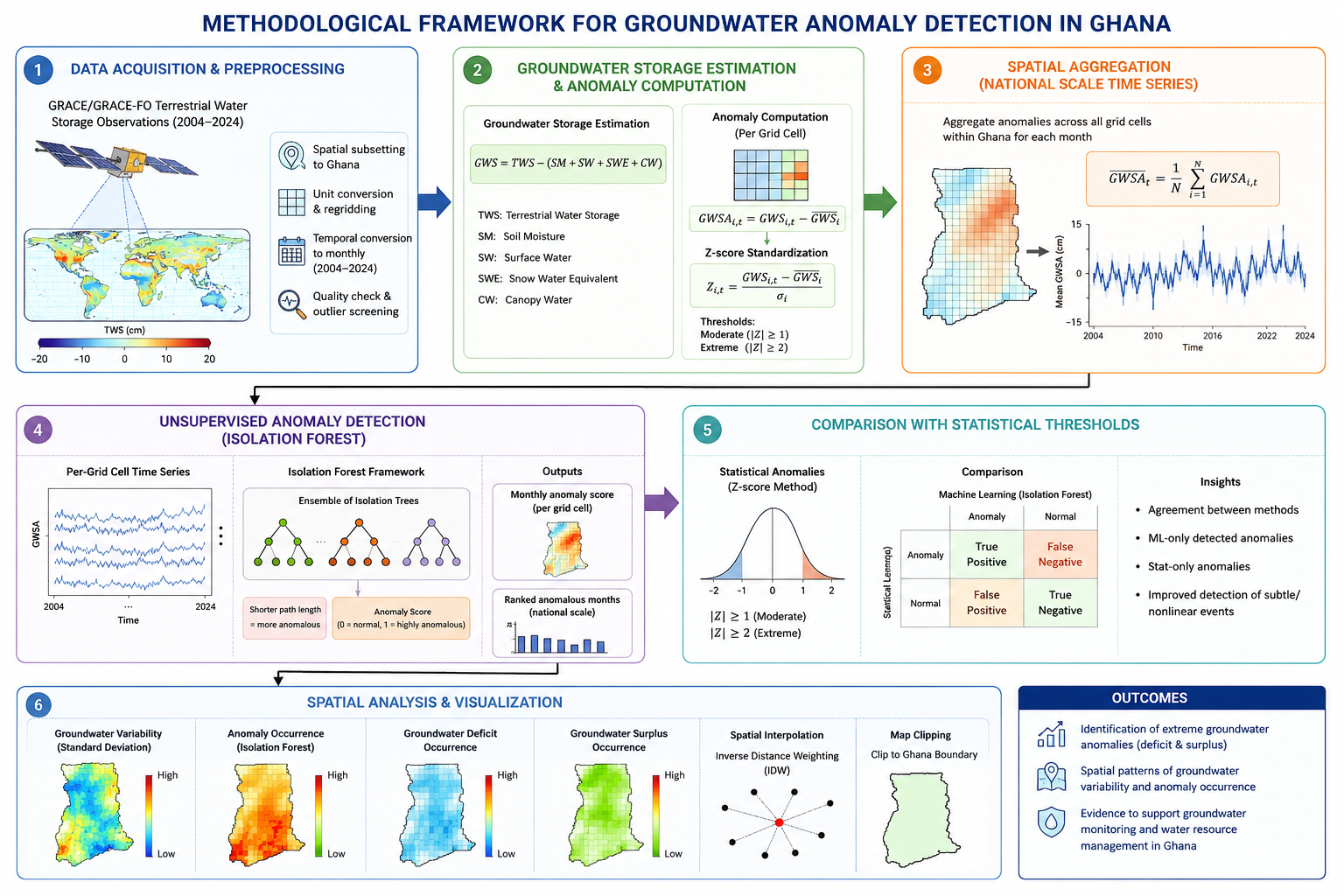} %
	\end{minipage}
	\caption{Workflow illustrating the integrated framework for groundwater anomaly estimation, unsupervised anomaly detection, statistical comparison, and spatial analysis across Ghana.}
	\label{fig:workflow}
\end{figure*}

\subsection{Spatial Analysis}

Spatial characteristics of groundwater variability and anomaly occurrence were investigated using grid-level statistical metrics computed over the full study period. These metrics included the spatial standard deviation of groundwater storage, anomaly occurrence frequency, groundwater deficit anomaly frequency, and groundwater surplus anomaly frequency.

Spatial representations of these metrics were generated using inverse distance weighting (IDW) interpolation to produce spatially smoothed maps of groundwater variability across Ghana. 
IDW was therefore used primarily as a visualization and spatial representation procedure and should not be interpreted as increasing the intrinsic spatial resolution of the GRACE observations
All spatial outputs were clipped to the Ghana boundary to ensure geographic consistency and improve interpretation of regional groundwater anomaly patterns.

The methodological workflow used in this study is summarized in Fig.~\ref{fig:workflow}.

\section{Results and Discussion} \label{sec:R4}

\begin{figure*}
	\begin{minipage}{\linewidth}
		\centering
		\includegraphics[width=\textwidth]{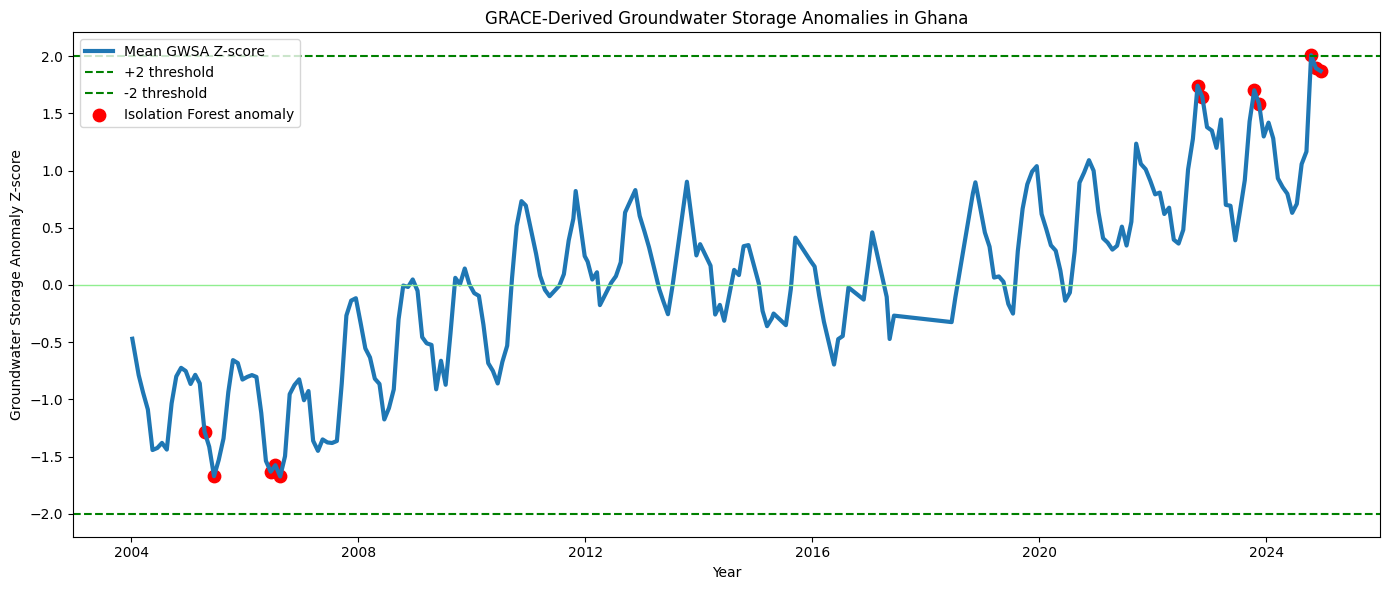}
	\end{minipage}
	\caption{Temporal variation of GRACE-derived groundwater storage anomalies (GWSA) in Ghana from 2004 to 2024 expressed as Z-scores. The blue line represents the spatially averaged groundwater anomaly, dashed horizontal lines indicate moderate ($\pm1$) and extreme ($\pm2$) anomaly thresholds, and red markers denote anomalous months identified using the ensemble-based anomaly detection framework.}
\label{fig:Time_series.png}
\end{figure*}

Figure~\ref{fig:Time_series.png} presents the temporal evolution of GRACE-derived groundwater storage anomalies in Ghana from 2004 to 2024. The time series exhibits pronounced seasonal oscillations superimposed on substantial interannual variability. Negative anomalies dominate the early portion of the record, particularly between 2004 and 2009, where several observations fall below the moderate anomaly threshold and approach extreme negative conditions. This period is followed by a transition toward more variable groundwater conditions between approximately 2010 and 2017, during which anomalies fluctuate around the long-term mean.

From 2018 onward, the groundwater anomaly series shows a gradual shift toward predominantly positive anomalies, with several observations exceeding the moderate threshold and approaching extreme positive conditions. These variations suggest substantial temporal changes in groundwater storage behavior over the study period and may reflect broader hydroclimatic variability influencing recharge and storage processes across Ghana.

The detected anomalous months, highlighted by the red markers in Fig.~\ref{fig:Time_series.png}, occur during both groundwater deficit and surplus phases, indicating that the ensemble-based detection framework captures deviations across different hydrological regimes rather than focusing exclusively on extreme deficits. A total of 12 anomalous months were identified using the anomaly detection framework, comprising 5 groundwater deficit events and 7 groundwater surplus events. Despite the larger number of surplus anomalies, the most intense anomalous events were associated with strong groundwater deficit conditions observed during 2006 and 2007.

\begin{figure*}[!t]
\centering

\begin{subfigure}[b]{0.32\textwidth}
    \centering
    \includegraphics[width=\linewidth]{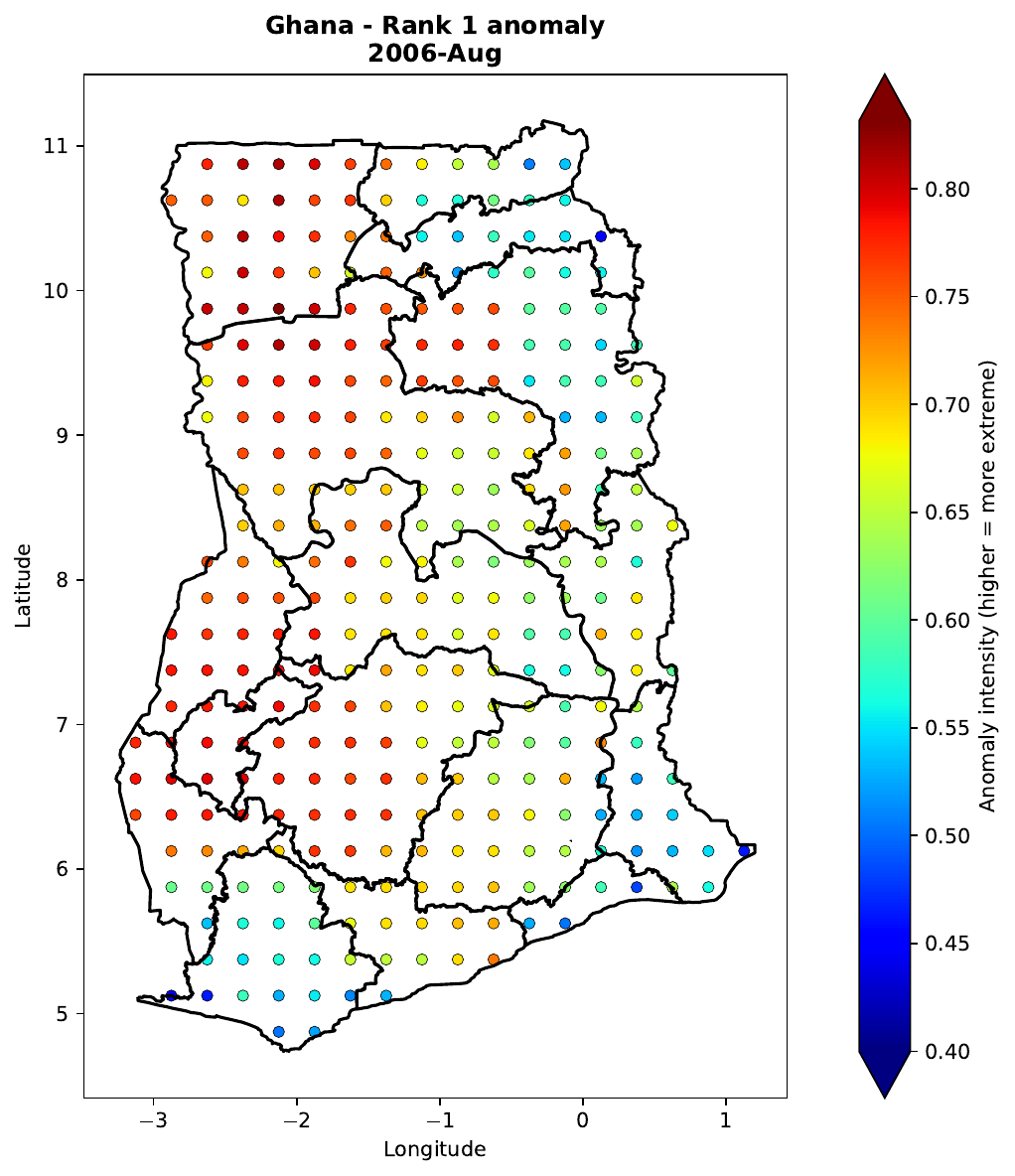}
    \caption{Aug 2006 (Rank 1)}
    \label{fig:rank1}
\end{subfigure}
\hfill
\begin{subfigure}[b]{0.32\textwidth}
    \centering
    \includegraphics[width=\linewidth]{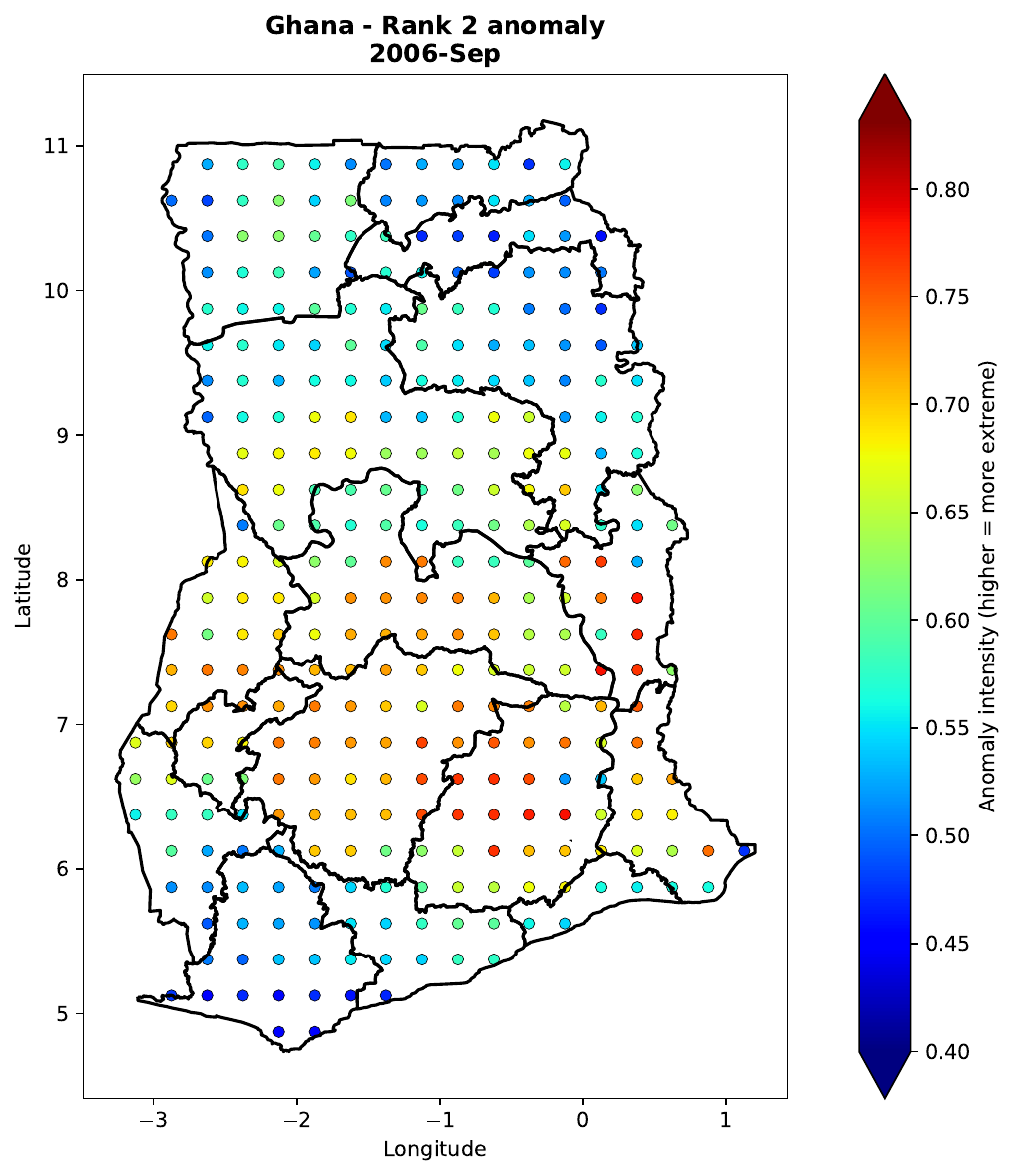}
    \caption{Sep 2006 (Rank 2)}
    \label{fig:rank2}
\end{subfigure}
\hfill
\begin{subfigure}[b]{0.32\textwidth}
    \centering
    \includegraphics[width=\linewidth]{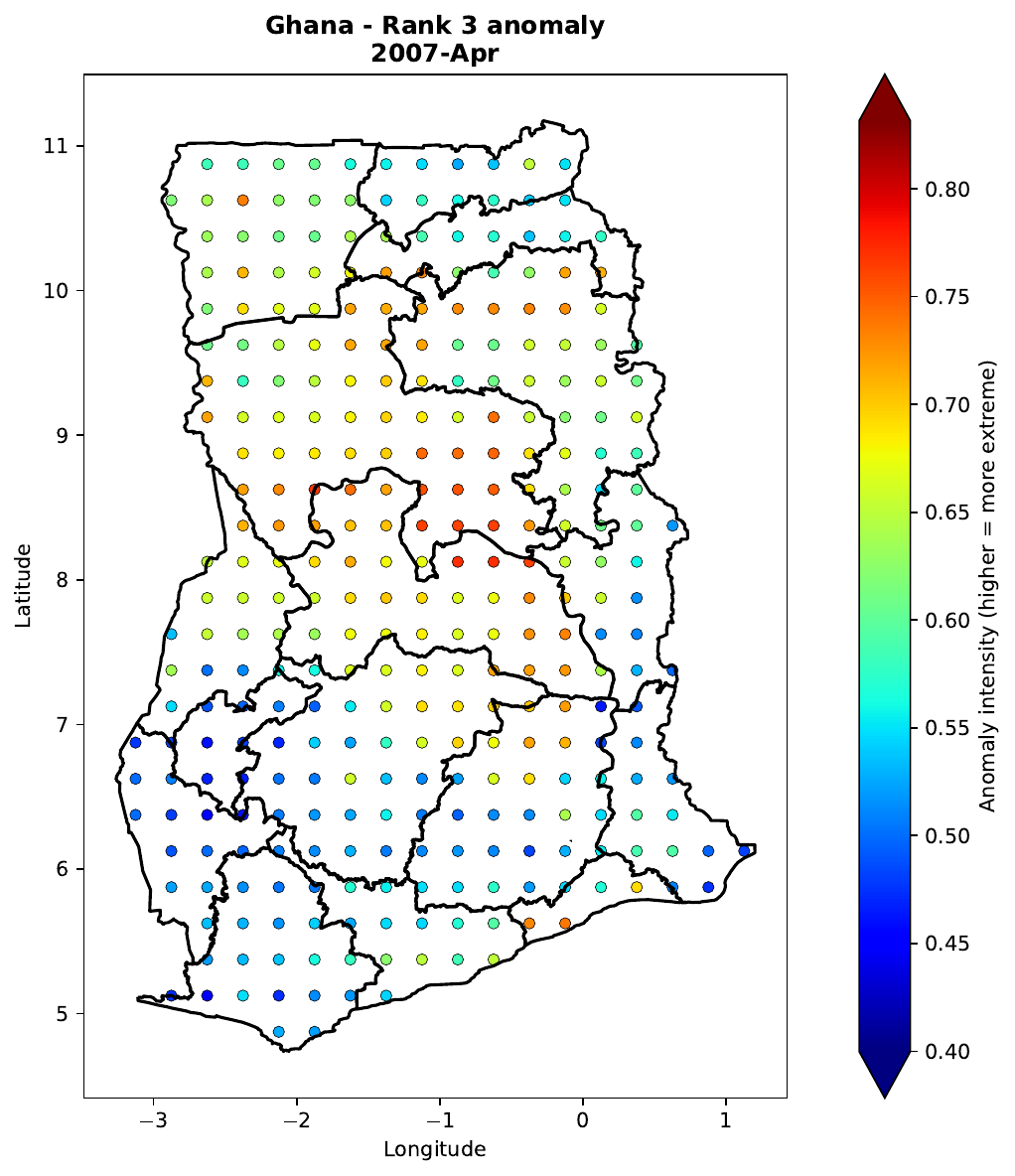}
    \caption{Apr 2007 (Rank 3)}
    \label{fig:rank3}
\end{subfigure}
\caption{Spatial distribution of the three most extreme groundwater storage anomalies in Ghana identified using the Isolation Forest algorithm: (a) August 2006 (Rank 1), (b) September 2006 (Rank 2), and (c) April 2007 (Rank 3). Colors represent anomaly intensity, with higher values indicating more extreme deviations.}
\label{fig:Anomaly_rank}

\end{figure*}

Figure~\ref{fig:Anomaly_rank} illustrates the spatial distribution of the three highest-ranked anomalous events identified from the monthly anomaly score ranking. The August 2006 and September 2006 events exhibit relatively widespread negative groundwater conditions across large portions of the country, whereas the April 2007 event displays greater spatial heterogeneity. These patterns indicate that extreme groundwater anomalies in Ghana may occur at both broad regional scales and localized spatial scales, depending on the prevailing hydrological conditions.


\begin{figure*}
	\centering
	\includegraphics[width=\textwidth]{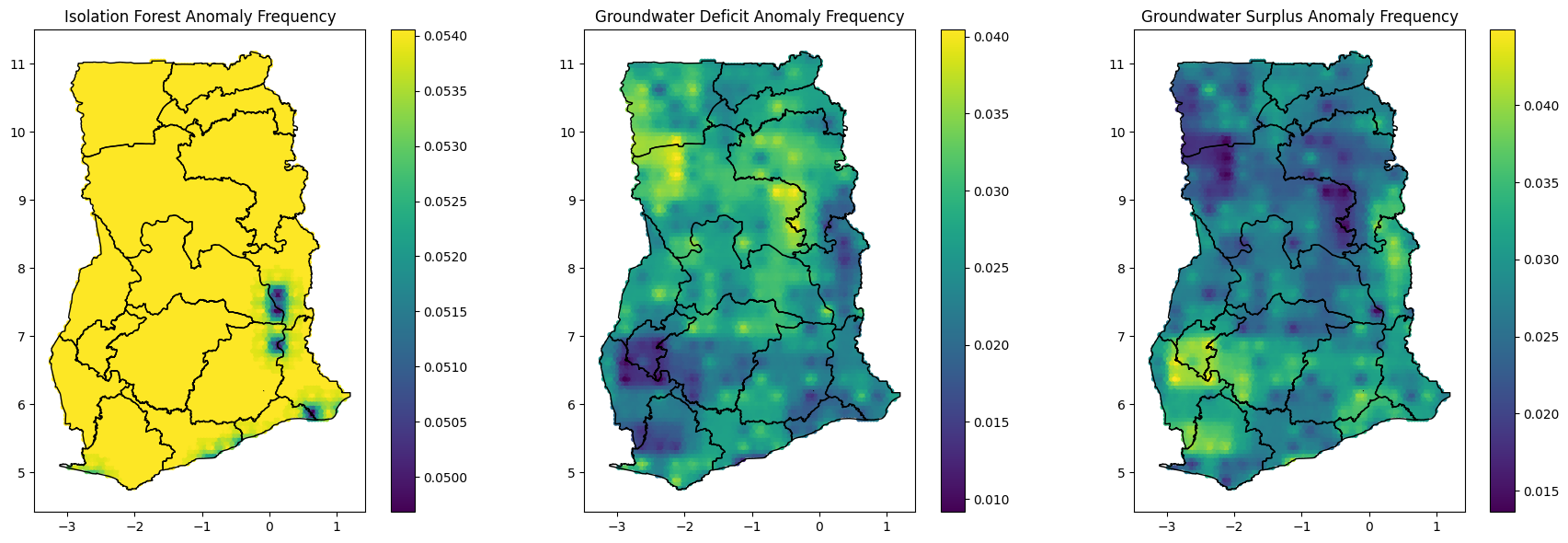}
	\caption{Spatial distribution of groundwater anomaly occurrence across Ghana for the period 2004--2024, showing total anomaly frequency (left), groundwater deficit frequency (middle), and groundwater surplus frequency (right) derived from the ensemble-based anomaly detection framework.}
	\label{fig:Spatial_freq}
\end{figure*}

The spatial distribution of groundwater anomaly occurrence frequencies is presented in Fig.~\ref{fig:Spatial_freq}. The overall anomaly occurrence pattern remains relatively spatially uniform because the anomaly proportion was constrained by the contamination parameter specified in the ensemble framework. However, separating anomaly occurrence into groundwater deficit and surplus categories reveals more distinct regional patterns.

Groundwater deficit anomalies occur more frequently across northern and central Ghana, whereas surplus anomalies are more pronounced in southern and southwestern regions. These spatial contrasts are broadly consistent with the hydroclimatic and geological variability across the country. Northern Ghana is characterized by longer dry periods and comparatively lower annual rainfall, while southern Ghana experiences wetter climatic conditions and more sustained recharge potential. In addition, fractured and weathered geological formations in parts of southern Ghana may facilitate relatively rapid recharge dynamics and increased temporal groundwater variability.

\begin{figure}
	\centering
	\includegraphics[width=0.5\textwidth]{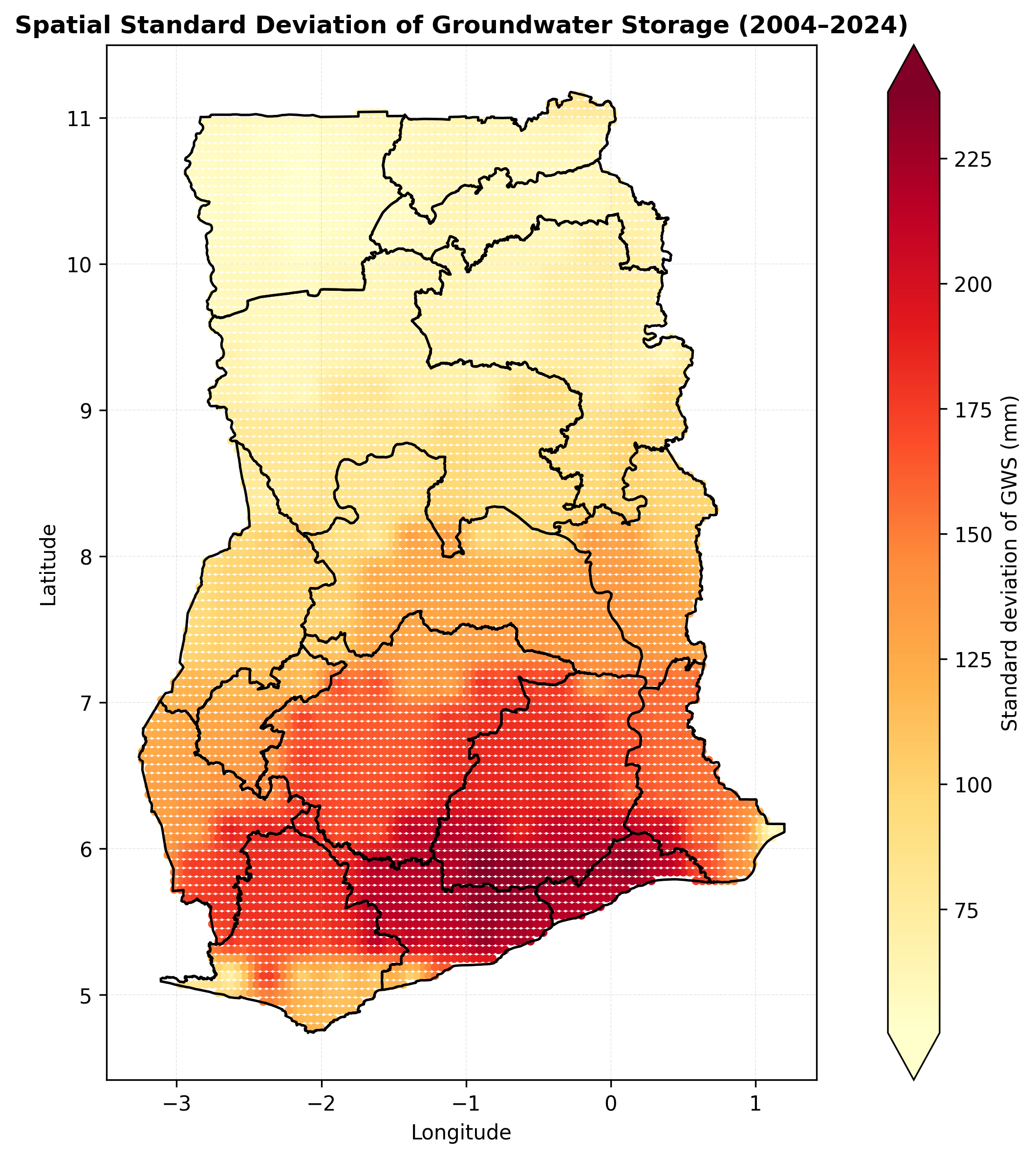}
	\caption{Spatial distribution of temporal groundwater storage variability across Ghana derived from GRACE observations for the period 2004--2024. Higher values indicate regions exhibiting stronger month-to-month fluctuations in groundwater storage conditions.}
	\label{fig:Storage_Var}
\end{figure}

\begin{figure}
	\centering
	\includegraphics[width=0.5\textwidth]{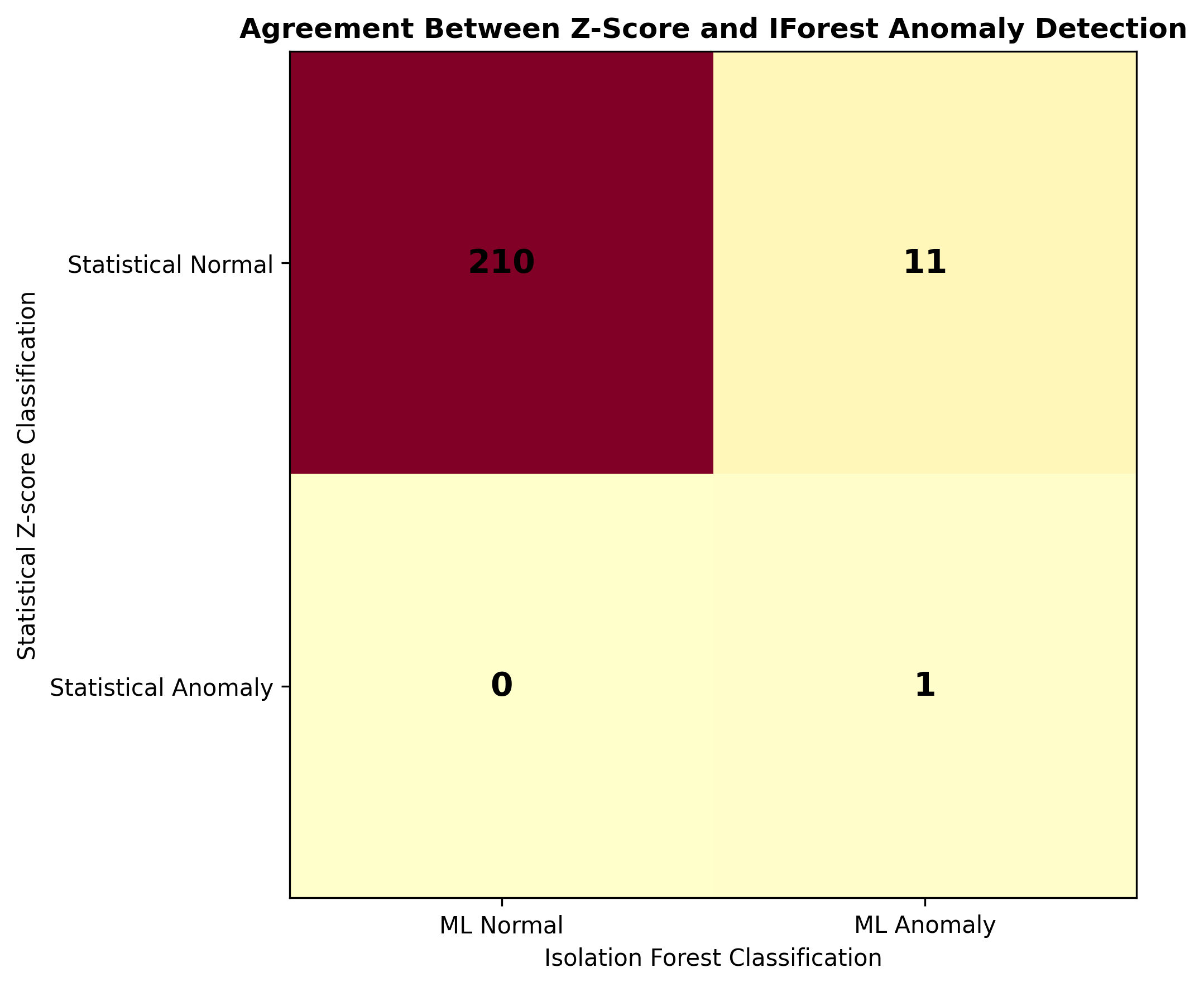}
	\caption{Confusion matrix comparing anomaly classifications obtained from the standardized Z-score method and the ensemble-based anomaly detection framework for groundwater storage anomalies in Ghana during 2004--2024.}	\label{fig:Std_confusion}
\end{figure}

Figure~\ref{fig:Storage_Var} shows that groundwater storage variability is spatially heterogeneous across Ghana, with stronger temporal fluctuations observed in central and southwestern regions and comparatively lower variability across parts of northern Ghana. The confusion matrix in Fig.~\ref{fig:Std_confusion} further indicates that the ensemble-based anomaly detection framework and the standardized Z-score method exhibit similar behavior in identifying normal groundwater conditions, while the machine learning approach additionally captures anomalous patterns that do not necessarily exceed fixed statistical thresholds. These differences suggest the presence of subtle or nonlinear groundwater variations embedded within the spatiotemporal anomaly structure. The comparison should therefore be interpreted as an assessment of methodological correspondence and anomaly sensitivity rather than a validation of one method against an absolute reference standard.

Overall, the results presented in Figures~\ref{fig:Time_series.png}--\ref{fig:Std_confusion} reveal that groundwater storage dynamics in Ghana are strongly influenced by the combined effects of climatic variability, recharge processes, and regional hydrogeological conditions. The temporal anomaly evolution indicates alternating phases of prolonged groundwater deficits and episodic surplus conditions, reflecting substantial interannual variability in groundwater storage behavior. Spatially, the observed north--south contrast in anomaly occurrence and groundwater variability is broadly consistent with Ghana’s hydroclimatic zonation and geological setting. Northern Ghana, characterized by longer dry periods and relatively lower recharge potential, exhibits more persistent deficit conditions and reduced temporal variability. In contrast, southern and southwestern regions display stronger groundwater fluctuations and more frequent surplus anomalies, potentially associated with wetter climatic conditions and enhanced recharge dynamics within weathered and fractured aquifer systems. The ranked anomaly maps further demonstrate that extreme groundwater events may occur at both regional and localized scales, emphasizing the spatial complexity of groundwater responses to hydroclimatic forcing. Collectively, these findings highlight the importance of integrating GRACE observations with data-driven anomaly detection techniques to improve understanding of groundwater variability and hydrological extremes in data-scarce environments.

\paragraph*{Limitations and Recommendations}
An important consideration in interpreting the spatial results is the coarse effective spatial resolution of GRACE observations (300--400 km). Given the relatively small size of Ghana and the heterogeneous distribution of its geological formations, aquifer systems, rainfall regimes, and groundwater-use pressures, groundwater variations occurring at local or sub-regional scales may be averaged within individual GRACE footprints. Consequently, the spatial patterns shown in Figures~\ref{fig:Anomaly_rank}--\ref{fig:Storage_Var} should not be interpreted as resolving individual aquifers or localized groundwater processes. Rather, they represent broad-scale patterns in the GRACE-derived groundwater storage signal. The apparent north–south contrasts in deficit and surplus anomaly frequencies are therefore more appropriately interpreted as regional-scale differences that are broadly consistent with Ghana's hydroclimatic and geological gradients. Similarly, the spatial interpolation applied to the GRACE-derived grid data improves visualization but does not add independent spatial information beyond the effective resolution of the satellite observations. Future studies combining GRACE/GRACE-FO with groundwater-level measurements, hydrological modelling, or higher-resolution remote-sensing products could provide more detailed characterization of local groundwater variability.

Overall, the results demonstrate that groundwater variability in Ghana exhibits substantial temporal and spatial heterogeneity, with both prolonged deficit periods and episodic surplus conditions contributing to anomalous groundwater behavior. The integration of GRACE observations with ensemble-based anomaly detection provides additional insight into groundwater variability beyond conventional threshold-based analysis and highlights the potential of unsupervised learning techniques for groundwater monitoring in data-scarce environments.




\section{Conclusion}
 
This study investigated groundwater storage anomalies in Ghana using GRACE-derived observations integrated with statistical and unsupervised machine learning approaches. The results revealed substantial temporal and spatial variability in groundwater storage between 2004 and 2024, characterized by prolonged deficit periods, episodic surplus conditions, and regionally varying anomaly occurrence patterns. Deficit anomalies were more pronounced across northern Ghana, whereas surplus anomalies and stronger groundwater variability were observed predominantly in southern and southwestern regions. The detected anomaly patterns highlight the influence of hydroclimatic variability and hydrogeological conditions on groundwater dynamics across the country.

The ensemble-based anomaly detection framework successfully identified both extreme and subtle groundwater anomalies beyond conventional threshold-based analysis, demonstrating the potential of unsupervised learning techniques for groundwater monitoring in data-scarce environments. The comparative analysis further showed that machine learning and statistical approaches provide complementary perspectives for characterizing groundwater variability. Overall, the integration of GRACE observations with data-driven anomaly detection offers a practical framework for regional groundwater assessment and may support improved monitoring of hydrological extremes and groundwater resource management in Ghana and similar hydroclimatic regions.
 
\section*{Acknowledgment}
The authors thank the anonymous reviewers for their constructive comments, which helped improve the clarity and quality of this manuscript. Prof.~Atemkeng also acknowledges financial support from Rhodes University, South Africa, towards the publication of this work.

\bibliographystyle{IEEEtran}
\bibliography{refs}

\end{document}